\documentclass[iop]{emulateapj}

\usepackage{amssymb}                    
\usepackage{graphics,graphicx}

\shorttitle{Statistical study of solar dimmings using CoDiT}
\shortauthors{Krista and Reinard}

\begin{document}

\title{Statistical study of solar dimmings using CoDiT}

\author{Larisza D. Krista$^{1}$ and Alysha A. Reinard$^{1,2}$}
\affil{$^{1}$University of Colorado/Cooperative Institute for Research in Environmental Sciences, Boulder, CO 80205, USA, \\ $^{2}$National Oceanic and Atmospheric Administration/Space Weather Prediction Center, Boulder, CO 80305, USA\\Corresponding author: \url{larisza.krista@colorado.edu} }

\begin{abstract}

We present the results from analyzing the physical and morphological properties of 154 dimmings (transient coronal holes) and the associated flares and coronal mass ejections (CMEs). Each dimming in our 2013 catalogue was processed with the semi-automated Coronal Dimming Tracker (CoDiT) using Solar Dynamics Observatory (SDO) AIA 193 \AA\ observations and HMI magnetograms. Instead of the typically used difference images, we used our coronal hole detection algorithm to detect transient dark regions ``directly'' in extreme ultraviolet (EUV) images. This allowed us to study dimmings as the footpoints of CMEs --- in contrast with the larger, diffuse dimmings seen in difference images, that represent the projected view of the rising, expanding plasma. Studying the footpoint-dimming morphology allowed us to better understand the CME structure in the low corona. Comparing the physical properties of dimmings, flares and CMEs we were also able to identify relationships between the different parts of this complex eruptive phenomenon. We found that larger dimmings are longer-lived, suggesting that it takes longer to ``close down'' large open magnetic regions. Also, during their growth phase, smaller dimmings acquire a higher magnetic flux imbalance (become more unipolar) than larger dimmings. Furthermore, we found that the EUV intensity of dimmings (indicative of local electron density) correlates with how much plasma was removed and how energetic the eruption was. Studying the morphology of dimmings (single, double, fragmented) also helped us identify different configurations of the quasi-open magnetic field.

\end{abstract}


\section{Introduction}

Coronal dimmings (or ``transient coronal holes'') are detected in the solar corona as areas of reduced emission in EUV and X-ray images. The formation of dimmings is generally explained by two governing theories. The prevailing hypothesis is that dimmings are caused by density depletion during a solar eruption. During the eruption plasma is removed and the previously closed magnetic field lines are dragged out, creating quasi-open magnetic field lines \citep{Thompson00, Harrison03, Zhang07}. The other hypothesis suggests that some secondary temperature variations could also be responsible for the formation of dimmings \citep{Thompson98, Chertok03}. However, the timescale at which dimmings form is much faster than the corresponding conductive and radiative cooling times \citep{Hudson96}. Furthermore, the simultaneous detection of dimmings at several different wavelengths in the same location suggests that temperature change alone cannot explain the appearance of dimmings \citep{Zarro99}. \citet{Harra07} have also found strong Doppler blue-shifted plasma outflows in dimmings. These studies collectively indicate that the primary cause for dimmings is density depletion rather than a temperature effect.

\subsection{Dimmings: defining the difference between observations and the physical phenomenon}
It is generally accepted in the solar physics community that dimmings are observed due to plasma evacuation. Nevertheless, the dimmings that are discussed in the literature often refer to different manifestations of the same phenomena \citep{Mason14}. Intensity depletions that are observed in different ways might be visualizing different parts of a dimming or a feature that somehow resembles a dimming but does not necessarily fit the above definition. It is of fundamental importance to recognize the difference between the identified phenomena and how they relate to the physical model. One of the most commonly accepted definitions of dimmings in the solar community is based on a physical model \citep{Thompson00, Attrill08} where the dimmings are the footpoints of an erupting loop structure (e.g. CME). In this model the dimming is defined as the region in the low corona where the plasma has been evacuated and therefore the intensity drops visibly in EUV wavelengths. These plasma outflows have been detected by \citet{Harra01} and \citet{ Tian12}. The dimming region remains dark because it is open magnetically, and any plasma that would ordinarily refill the region is able to escape freely along the open magnetic field lines \citep{Sterling97}. This region eventually closes down as the plasmoid in the erupted structure disconnects and the open field lines get pinched to create a closed magnetic field where the plasma can accumulate to allow for the dimming region to disappear (brighten in the EUV image). This definition, we believe, is a complete description of dimmings -- it describes the 3D structure of the dimming, it relates it to the eruptive phenomena and the observed feature in low corona observations. Note that we intentionally made an adjustment to the accepted definition: rather than assuming all dimmings are related to CMEs, the less specific ``eruptive phenomena'' term is used. This is because many dimmings have been detected without an observable CME. This could be due to the fact that the CME was oriented in an unfavorable way to observe or too small to be seen in coronograph images, or alternatively, the eruption was a type of ``micro-CME" (smaller than a regular CME, with the possibility of plasma falling back to the Sun).

Until recently, the most common way of observing dimmings was through difference images. Both base and running difference images are very good at highlighting dimmings, as the plasma evacuates from the region where the dimming forms. A base difference image results from the subtraction of a pre-eruption image from a post-eruption image. Because the post-eruption image shows the dimming region with lower intensity, the subtraction enhances this region and therefore the dimming appears black. This subtraction can be done for a sequence of images, which shows the gradual evacuation of the plasma from the footpoint region. However, this process not only shows the footpoints of the erupted loop structure, but the displacement of the plasma throughout the corona until the density drops to a level where it is no longer detectable. For this reason, the difference images show large, wide-spread, diffuse dimming regions where the boundaries are challenging to specify. Running difference imaging is a similar method to base difference, except it is the n-1 image that is subtracted from the n-th image. It similarly shows a diffuse widespread structure for the dimming. 

Difference imaging is a very good method to identify even faint dimmings, as the subtraction significantly enhances the dimming. However, it is important to understand that when using this method it is not only the footpoints of the erupting loop structure that are being imaged, but also the whole ``bubble'' structure of the rising plasma. From here on we will refer to this type of dimmings as ``projection-dimmings'', since in these phenomena we mostly observe the 2D-projected image of the rising plasma of the CME. This means that when studying the relationship between CMEs and the projection-dimmings, there is a possibility of self-correlation. 

A more recently developed method of dimming detection is the {\it direct} imaging of dimmings in EUV wavelength images \citep{Krista13}. Here, only the loop footpoints are observed; therefore, we will refer to these dimmings as ``footpoint-dimmings'' from here on. These dimmings show the region in the low corona where the plasma was ejected during the eruption. The observations are best done close to the disk center where line-of-sight (LOS) effects are minimal. By studying the footpoint-dimmings, we compare footpoint properties alone with CME properties, rather than an amalgam of the footpoint and the CME plasma (as it appears in the difference images) with the CME properties.

\subsection{Previous studies}
Most studies on dimmings and the related eruptive phenomena are derived only from a handful of events. One of the first large statistical analyses was carried out by \citet{Reinard08} on 96 {\it on-disk} dimmings detected between 1998--2000. In this study \citeauthor{Reinard08} found that the dimming lifetimes were between 3 to 12 hours and they typically appeared in the belt of active regions (20-50 degrees latitude). In their next study \citet{Reinard09} found that dimming-associated CMEs have higher speeds and are more likely to be accompanied by flares. Furthermore, it was found that CME speed is weakly correlated with dimming size, and the magnetic flux in dimming regions does not correlate with the CME speeds. 

\citet{Bewsher08} also analyzed a large dataset of 200 {\it off-disk} dimming events from 1998 to 2005 using EUV data. Note, however, that this work is different from the rest of the studies discussed here: the authors detected dimmings at the solar limb and the dimmings were analyzed using spectroscopic data. They found that up to 55\% of the dimming events were associated with CMEs and up to 84\% of CMEs in their dataset could be tracked back to dimmings. \citet{Bewsher08} also note that one emission line may not be sufficient for associating dimming regions with CMEs.

Studying one event \citet{Zhukov04} found that the transient coronal hole (what we call footpoint-dimming -- seen in direct imaging) is a third of the size of the dimming (what we call projection-dimming -- seen in the difference images). \citet{Zhukov04} also estimated that half of the CME mass originates from the transient coronal hole (footpoint-dimming).

The goal of our work is to improve the precision of dimming detection by limiting the identification to footpoint-dimmings and by automating the effort as much as possible. Since the footpoint-dimmings we study have clearer boundaries, we are able to follow their morphological changes and track their evolution carefully. This careful identification method and our extensive dimming catalogue allow us to better understand the relationship between dimmings and CMEs. The details of how the detection is performed are discussed in Section \ref{method}. The catalogued dimmings were also linked to CMEs and flares to investigate their relationship (Section \ref{results}). We conclude our work by discussing the morphological types of dimmings and the identified dimming-CME-flare relationships that could be beneficial for space weather forecasting efforts (Section \ref{conclusions}).

\section{Observations and Data Analysis}
\label{method}

We detect dimming regions using 193 \AA\ observations taken by the Solar Dynamics Observatory (SDO) {\it Atmospheric Imaging Assembly} (AIA; \citeauthor{Lemen12} \citeyear{Lemen12}) instrument. This wavelength is commonly favored for open magnetic field region (dimming and CH) detection due to its high contrast between the dark open regions and the brighter quiet Sun (QS). A further advantage to using SDO/AIA data is its high spatial and temporal cadence, which allows us to study the morphology and temporal evolution of dimmings in more detail. In addition to the EUV images, we use the SDO {\it Helioseismic and Magnetic Imager} (HMI; \citeauthor{Scherrer12} \citeyear{Scherrer12}) magnetograms to determine the LOS photospheric magnetic field properties of dimmings (e.g. polarity, magnetic field strength). This is particularly important as dimmings are thought to be the footpoints of CME flux-ropes, and therefore could provide information on the link between dimmings and the magnetic configuration of CMEs. 

First, we have manually identified dimming regions using daily movies of AIA 193 \AA\ observations for the year 2013. The visual criteria for dimming identification was a region that was a transient, dark phenomenon, without clear obscuration effects from loop structures. In total, 115 dimming systems were identified visually. The manual identification was followed by the semi-automated detection using the Coronal Dimming Tracker algorithm (CoDiT; \citeauthor{Krista13},  \citeyear{Krista13}). Based on the visual dimming catalogue, CoDiT identified 154 dimmings (including fragments in the same dimming system). Out of these dimmings, 85 were ``single'' dimmings (one dominant dimming), 22 were ``double'' (with 2 main dimmings in the system) and 8 were "fragmented" (a dimming system with more than two dimmings). The morphological classification of the dimmings is necessary to study the dimming region as a whole. The classification of dimmings was done manually using the time of their best visibility. However, dimmings can split and merge during their lifetimes, therefore, the classification is most descriptive of the time of peak visibility. This does not influence our statistical analysis as dimming systems are treated as one - e.g. at any given time the area is the sum of the dimming components in the system. It is easiest to analyze a single dimming: here only one dimming is tracked from its emergence until it disappears. Fragmented and double dimmings have to be tracked as a system. This is done by identifying the fragments and tracking their evolution separately. It is important to note that in addition to studying dimming fragments as a system, we also allow the further fragmentation and merging of each individual dimming. This means that during the lifetime of each dimming (whether it is a single dimming or part of a dimming system), we track the dimming and identify any splitting or merging during its lifetime. This thorough analysis is necessary to precisely study the evolution of dimming properties (lifetime, area, magnetic properties, etc.).

The basis of the thresholding and tracking methods used in our present work are described in detail in \citet{Krista09, Krista11} and \citet{Krista13}. We have since made major improvements to the previous versions of our open field detection algorithms. These improvements are bringing us closer to fully automating our dimming detection and allowing us to study dimmings as a system rather than individual dimmings. 

We describe the most important steps of the detection and analysis here. The CoDiT image analysis starts with the calibration of the AIA data using the {\it aia\_prep} routine (see SolarSoft; \citeauthor{Freeland98} \citeyear{Freeland98}). The EUV disk image is first transformed to a Lambert cylindrical equal-area projection map. The Lambert maps aid the thresholding and tracking process, and the projection corrects for LOS distortions on spherical disk images. The dimmings are detected in the image using a local intensity thresholding method, which distinguishes low-intensity regions (e.g. dimmings or CHs) from higher-intensity QS regions. As CoDiT is currently semi-automated, the user defines the size and location of the region where the dimming is detected and the time when the dimming was well observed. This basic information about location and time is the only required manual input and is enough for the algorithm to then automatically detect the dimming and track it to the time of emergence by stepping back in time. The emergence is determined as the time when the dimming has reached a minimum area threshold. Once the emergence time is established, the tracking starts again to make sure all fragments of the dimming are identified and tracked over its lifetime until it disappears. For this continuously changing morphology, we developed a tool to track fragmenting and merging parts of the dimming system. With each AIA detection of the dimming a magnetogram is found close to the AIA observation time, and the dimming contour is overlaid to determine the magnetic properties of the dimming. From the AIA images we determine a plethora of properties: the dimming boundary threshold; the mean EUV intensity, area and location at each instance of dimming observation (every 4 minutes); the change of intensity from 1 hour before the dimming emergence to the time of maximum area (using the maximum area dimming contour); the ascend time (from emergence to the time of peak area); the descend time (from the peak area time to the time of disappearance); total lifetime of the dimming; and the EUV intensity of dark cores within the dimming (we assume that the lower 10\% intensity regions within the dimmings are flux concentrations). From the magnetograms we can determine the mean magnetic field strength and overall polarity at each observation, the magnetic field skewness (or flux imbalance) at the maximum area time, the change in the magnetic field strength and skewness from 1 hr before the dimming emergence to the maximum area time. This is done by overlaying the dimming boundaries detected in the AIA images on the corresponding HMI magnetograms. Although dimmings are coronal phenomena, they are essentially ``holes'' in the corona, and hence it is reasonable to gather information on the photospheric magnetic field in the area traced out by the dimming boundary.

We also link the dimming properties to CME and flare properties to better understand the connection between the low-corona signature of the eruption to the eruptive counterpart on- and off-disk. We used the National Oceanic and Atmospheric Administration (NOAA) Geostationary Operational Environmental Satellite (GOES) flare list to link flares to the dimmings in our catalogue. We visually inspected EUV images to identify if the dimmings and flares occurred in a nearby location -- e.g. at the same AR. If no AR was associated, physical proximity helped to make the link. If the flare was not visible at all, then we relied on the timing of the dimming emergence and the flare occurrence. Amongst the dimming-flare associations we found that 90 percent of all flares occurred within $\pm$ 2 hours of the dimming emergence. In order to link dimmings and CMEs we used the Solar and Heliospheric Observatory (SOHO) Large Angle and Spectrometric Coronagraph (LASCO) CME catalog (for more information see: \href{http://cdaw.gsfc.nasa.gov/CME\_list/catalog\_description.htm}{http://cdaw.gsfc.nasa.gov/CME\_list/catalog\_description.htm}). For each dimming we looked for a CME and visually identified the connection using the difference image movies of the CMEs (e.g., the dimming location on the Sun gives an indication of where we might see the CME). The difference image movies show the solar disk and also part of the corona, and often it is possible to visually identify the dimmings and the corresponding CME in the same movie. We note that the dimmings we observed were {\it on-disk} and within $\pm$70 degrees longitude and hence the resulting CMEs were not in the plane of sky. It is understood, that the LASCO CME catalogue assumes all CMEs to be in the plane of sky for the calculated properties -- and therefore the errors are lowest when the CME is in fact in the plane of sky. 

Having completed our catalogue we found that 85\% of CMEs occurred within $\pm$2 hours after the dimming emergence. For the CME times we used the date and time of their first appearance in the LASCO/C2 field of view. We found 64 CME-dimming and 41 flare-dimming links.


\section{Results}
\label{results}

Using the CoDiT algorithm we had a unique opportunity to objectively identify and analyze dimmings in a one-year period (2013). The resulting dimming catalogue was then used to statistically analyze the properties of dimmings and establish their morphological characteristics. In addition, we manually linked dimmings to their eruptive counterparts (CMEs and flares) and investigated their relationship. Studying the connection between the open magnetic field region and the observational evidence to the destabilization of the source region helps us better understand the eruptive process as a whole. We emphasize that the findings detailed below are indicative of correlations, but not necessarily causation. Also - associations between properties become more significant the closer the correlation coefficient is to 1. With values of 0.5 or lower there is a {\it possibility} of association. In Section \ref{conclusions} we further discuss how the correlation between the dimming, CME and flare properties might be indicative of a physical relationship.

\subsection{Statistical Analysis of Dimming Regions}

Analyzing the 2013 SDO/AIA 193 \AA\ images we identified 154 dimmings. The dimming properties were determined and statistically analyzed to better understand the general characteristics of footpoint-dimmings -- as shown in Table \ref{table:table1}. {\it Lifetime} is the time during which the dimming was identifiable using our algorithm. Ascend time ($\Delta t_{asc}$) is the time from the emergence to the time when the dimming reached its peak area, and the descend time ($\Delta t_{desc}$) is the time from peak-area to when the dimming disappeared. {\it Area} refers to the maximum area a dimming reached over its lifetime. $|\langle B_{LOS}\rangle|$ is the absolute value of the mean LOS magnetic field at {\it peak-time} (when the dimming reached its largest area), and $\langle I_{EUV} \rangle $ is the mean EUV intensity of the dimming determined at peak-time. 

\begin{table}[t]
\caption{Dimming properties}
\centering
\begin{tabular}{ c c c c c }
\hline \hline
Property & Mean & Median & Min & Max \\ \hline
Lifetime [hrs] &  9 & 5 & 1 & 138 \\
$\Delta t_{asc}$ [hrs] &  3 & 1 & 0.1 & 35 \\
$\Delta t_{desc}$ [hrs] &  6 & 3 & 0.3 & 117 \\
Area [$10^{3}$ Mm$^{2}$] &  4 & 2 & 0.3 & 19 \\
$|\langle B_{LOS}\rangle|$ [G] & 1 & 1 & 0 & 15 \\
$\langle I_{EUV} \rangle $ [DNs] & 128 & 128 & 64 & 172 \\
\end{tabular}
\label{table:table1}
\end{table}
\begin{figure}[!t]
\centerline{
\includegraphics[scale=0.35, trim=40 20 0 300, clip]{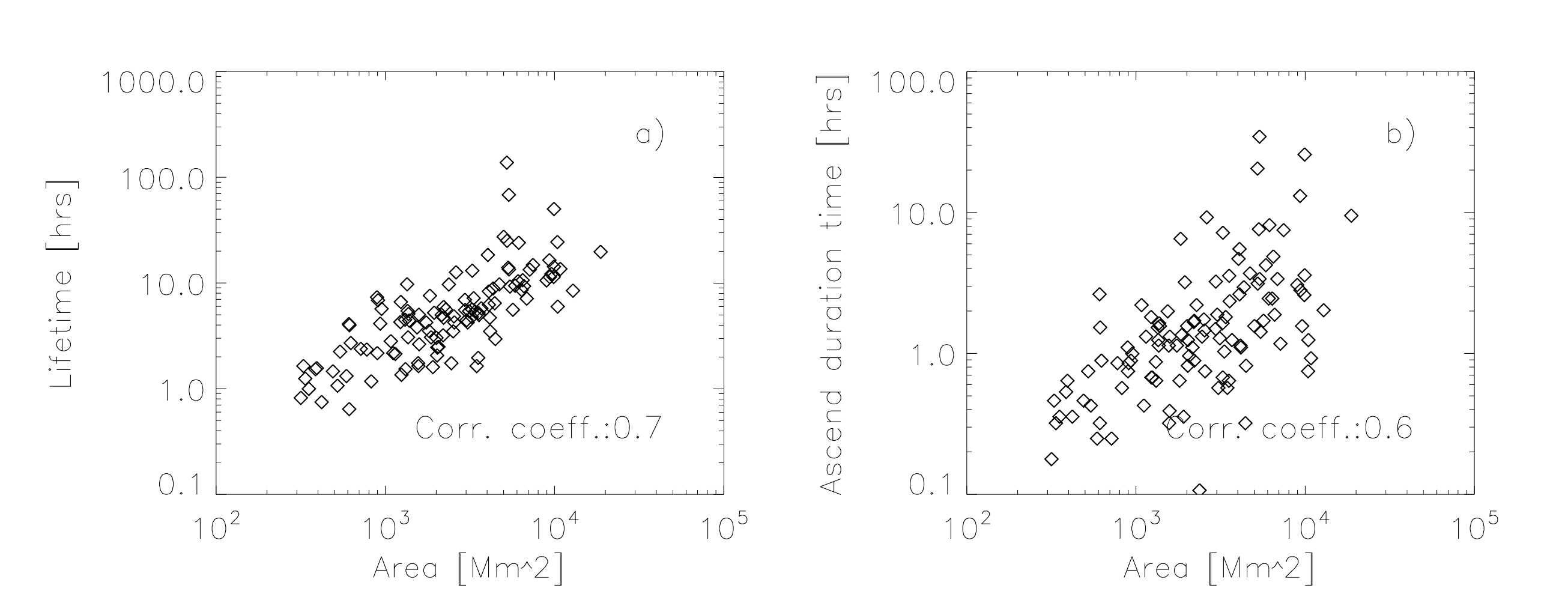}}
\centerline{
\includegraphics[scale=0.35, trim=40 20 0 300, clip]{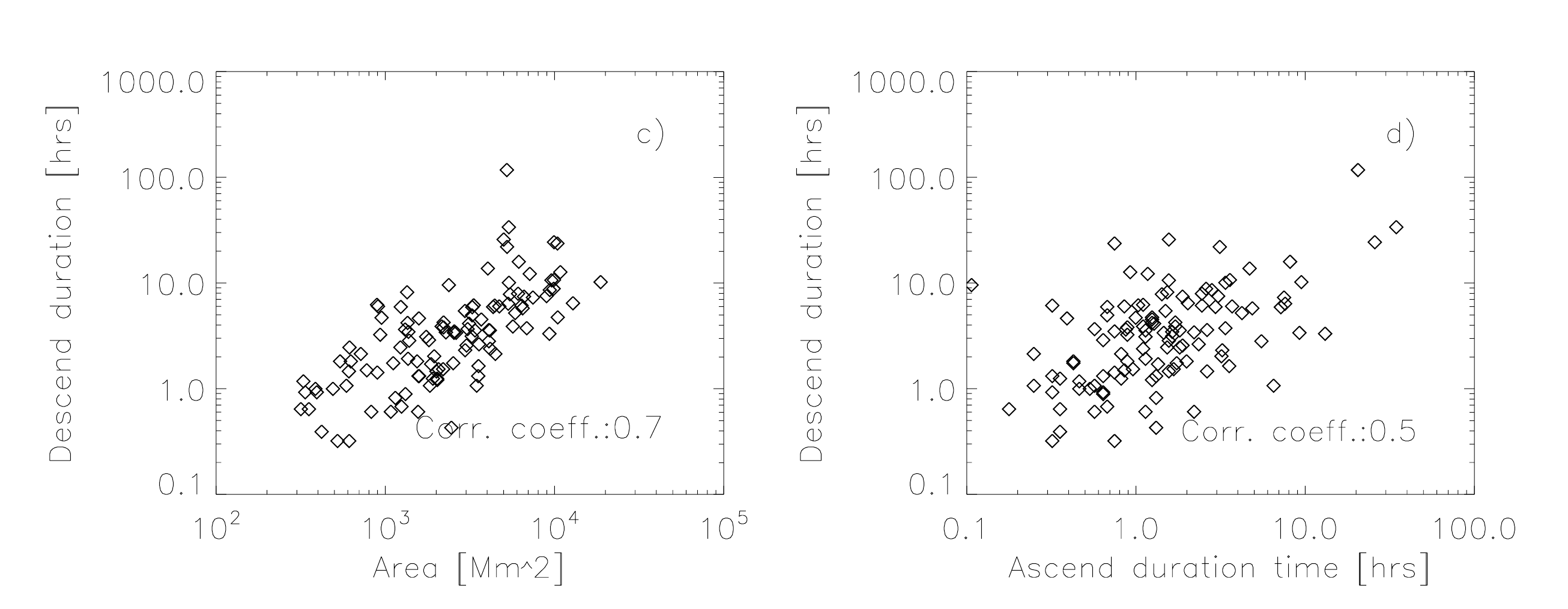}}
\centerline{
\includegraphics[scale=0.35, trim=40 20 0 300, clip]{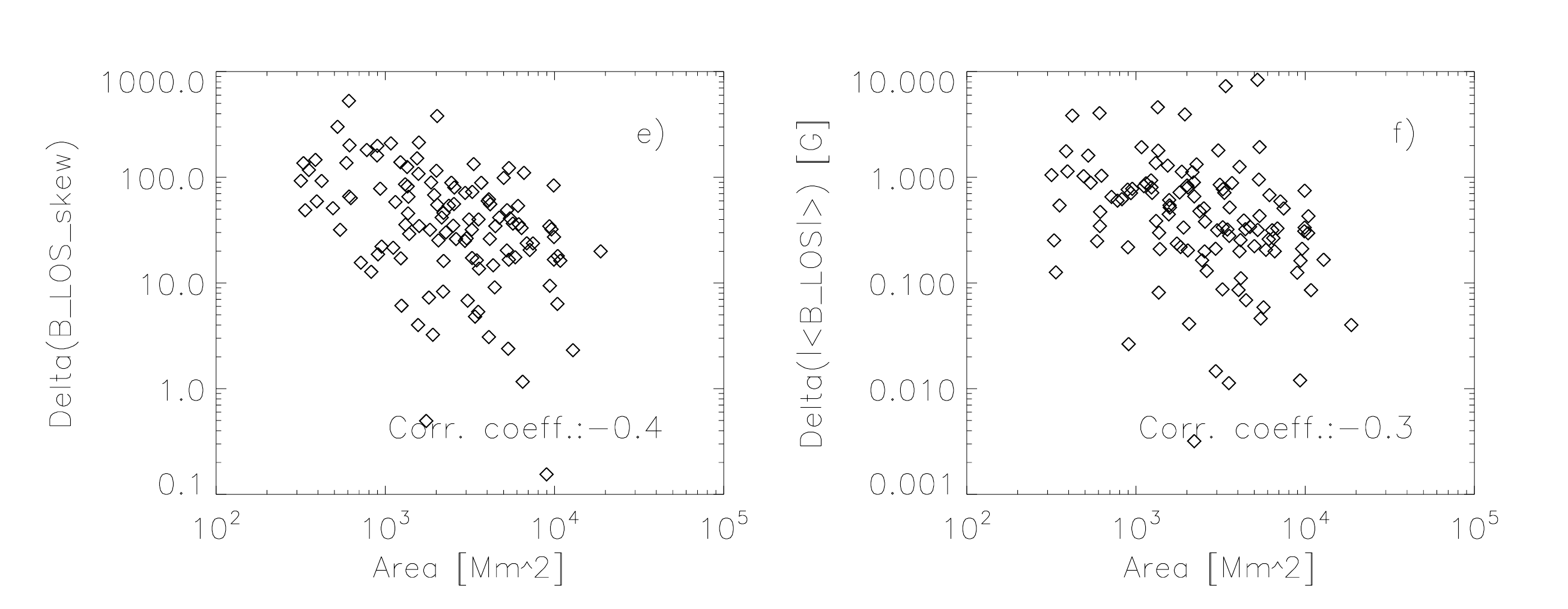}}
\caption{Scatterplots showing the positive correlation between dimming area and lifetime (Fig. 1a); dimming ascend (Fig. 1b) and descend-time (Fig. 1c); the correlation between dimming ascend and descend time (Fig 1d); the negative correlation between the dimming area and the change in LOS magnetic field strength skewness (Fig. 1e), and the change in the mean LOS magnetic field strength (Fig. 1f).}
\label{dim_scatter}
\end{figure} 

The dimmings in our catalogue were identified and studied between $\pm 70$ degrees longitude and $\pm 50$ degrees latitude. We chose not to include dimmings that were too close to the limb (further than 70 degrees latitude/longitude) to lower LOS effects in the detection, since the closer we get the the solar limb, the more obscuration there is in the observations. Our statistical analysis showed that the mean lifetime of dimmings is 9hrs, the ascend time is 3 hrs and the descend time is 6hrs. When analysing dimmings at their peak-time, we found that the mean area is $4\times 10^{3} Mm^{2}$. The absolute mean magnetic field was found to be 1.4 G (shown rounded down in the Table \ref{table:table1}), and 98\% of the observations show a mean magnetic field under 5G; therefore, the magnetic field in dimmings is typically weak. In comparison, the typical QS mean magnetic field strength is $\sim$1 G \citep{Lin99}. The mean EUV intensity of dimmings is 128 DNs with a standard deviation of 21 DNs. Therefore, 95\% of dimmings have a threshold intensity between 86--170 DNs for SDO/AIA 193 \AA\ observations.


We also looked at the correlation between different dimming properties (Figure \ref{dim_scatter} and Table \ref{table:table2}). The strongest correlation we found (0.75) was between the logarithmic values of the dimming area and lifetime, which clearly shows that large dimmings have a longer lifetime (Figure \ref{dim_scatter}a). We also found a correlation between the log values of the dimming area and the ascend and descend time (correlation coefficients were 0.6 and 0.7, respectively -- Figure \ref{dim_scatter} b \& c); consequently dimmings that have a longer ascend times are more likely to have a longer descend times. The latter was found to have a correlation coefficient of 0.5 (Figure \ref{dim_scatter}d). 

\begin{figure}[!t]
\centerline{
\includegraphics[scale=0.33, trim=30 120 0 120, clip]{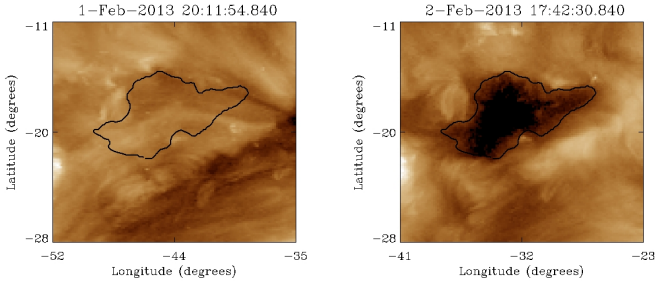}}
\caption{An example of a dimming location shown 1 hour before the emergence (left), and the same contour shown at the time of maximum area (right). The dimming shown was observed on 1-2 February 2013 using SDO/AIA 193 \AA\ images.}
\label{before_after}
\end{figure}

We also investigated the change of properties in dimmings over their evolution. Particularly, we were interested to analyze the same region pre-eruption and post-eruption. For this we have used the dimming contour at peak-time (when the area was maximum), determined the physical properties and then overlaid the same contour in the same region 1 hour before the dimming emerged. This meant comparing the physical properties of the the dimming with those of the pre-eruption dimming emergence site in the QS. An example of the pre- and post eruption dimming site is shown in Figure \ref{before_after}: on the right the dimming is seen at its peak-time and on the left the same region is shown one hour before the dimming emerged.

When studying the flux imbalance (skewness) change between pre-eruption and peak-time ($\Delta Skew(B_{LOS})$), we found that the log of the change in skewness anti-correlates with the log values of the dimming area (correlation coefficient -0.4, Figure \ref{dim_scatter}e). Therefore, larger dimmings show less change in their flux imbalance, than smaller dimmings. A similar anti-correlation (-0.3) was found between the log values of the dimming area and the change in the absolute mean LOS magnetic field $\Delta |\langle B_{LOS}\rangle|$ (Figure \ref{dim_scatter}f), suggesting that smaller dimmings experience more change in the magnetic field strength from pre-eruption to peak-time. 

\begin{table}[t]
\caption{Pearson correlation of dimming properties (all $lg$ values)}
\centering
\begin{tabular}{ c c c c c c }
\hline \hline
$lg$  & Lifetime & $\Delta Skew(B_{LOS})$ & $\Delta |\langle B_{LOS}\rangle|$ & $\Delta t_{asc}$ & $\Delta t_{desc}$ \\
& [hrs] &  &  [G]&  [hrs] &  [hrs] \\ \hline
Area &  0.7 & -0.4 & -0.3 & 0.6 & 0.7 \\
$[Mm^{2}]$ &  &  &  &  &  \\
\end{tabular}
\label{table:table2}
\end{table}


\subsection{Relationship between dimmings, flares and CMEs}

Where possible, the catalogued dimmings were linked to flares listed in the NOAA National Centers for Environmental Information (NCEI) GOES X-ray catalogue, and CMEs listed in NASA's Coordinated Data Analysis Workshop (CDAW) catalogue. We then analyzed the dimming properties (area, lifetime, EUV intensity, LOS magnetic field strength, and changes in properties from pre- to post-eruption) and compared them to the properties of the corresponding flares (duration, X-ray intensity) and CMEs (velocity, acceleration, mass, angular width, kinetic energy). The correlation coefficients are shown in Table \ref{table:table3}. The CME kinetic energy is shown as $E_{CME_{kin}}$, the angular width as $\phi_{CME}$, the mass as $m_{CME}$, the flare duration as $\Delta t_{flare}$ and the flare X-ray intensity as Flare $I_{xray}$.
The dimming mean EUV intensity at peak-time is shown as $\langle I_{EUVpeak} \rangle$, the change in the dimming mean intensity from pre-eruption to peak-time as $\Delta \langle I_{EUV} \rangle$ and the change in the mean value of the lowest 10\% intensity pixels as $\Delta \langle I_{low10\%} \rangle$. The significant correlation values are listed in Table \ref{table:table3}.

\begin{table}[t]
\caption{Pearson correlation of dimming, CME and flare properties}
\centering
\begin{tabular}{ c c c c c }
\hline \hline
  & $lg(E_{CME_{kin}})$ & $lg(m_{CME})$ & $lg(\Delta t_{flare})$ & $\phi_{CME}$ \\
    & [erg] & [g] & [s] & [Deg]\\ \hline
$\langle I_{EUVpeak}\rangle$ &  -0.4 & -0.4 & -0.3 & - \\
$\Delta \langle I_{EUV}\rangle$ & - & -0.4 & -0.5 & -0.4  \\
$\langle I_{EUVpre}\rangle$  & - & 0.5 & - & 0.4  \\
$\Delta \langle I_{low10\%}\rangle$ & -0.4 &  -0.4 & -0.5 & -0.4  \\
$[DNs]$ &  &   &  &  \\
\end{tabular}
\label{table:table3}
\end{table}



Furthermore, we found correlations between CME and flare log properties, listed in Table \ref{table:table4}. These findings confirm the widely known and accepted relationship between CMEs and flares. 


\begin{table}[t]
\caption{Pearson correlation of flare and CME properties (all $lg$ values)}
\centering
\begin{tabular}{ c c c c c c }
\hline \hline
$lg$  & $m_{CME}$ [g] & $\phi_{CME}$ [Deg] & $E_{CME_{kin}}$ [erg]  \\ \hline
$\Delta t_{flare}$ [s] &  0.5 & 0.4 & 0.4 \\
Flare $I_{xray}$ [$J/m^{2}$] &  0.4 & - & -  \\
\end{tabular}
\label{table:table4}
\end{table}

\section{Discussion and Conclusions}
\label{conclusions}

The work presented here investigates multiple aspects of dimmings using observational techniques. We determined the morphological characteristics of dimmings, their physical properties and their temporal evolution. Furthermore, we investigated the association between the properties of dimmings, flares and CMEs to help us better understand the relationship between the different parts of this complex eruptive process: the eruption, the open field created through the removal of plasma and extension of magnetic fields into the corona and beyond. We were particularly interested in the relationship between dimmings and the eruptive counterparts, since using dimmings for forecasting large geomagnetic storms would be highly beneficial. In our current catalogue we were able to link 64 dimmings with CMEs and 41 dimmings with flares.

\subsection{Relationship between dimmings, flares and CMEs}

When investigating dimming, flare and CME properties, we found several moderate and strong correlations. The most significant correlation we found (0.75) was between the logs of the dimming area and lifetime, which indicates that large dimmings have longer lifetimes. We suggest that when a large closed region is opened up through an eruption, it takes longer time to close a larger region down (as opposed to a smaller region) and fill it up with plasma once again in a stable magnetic configuration. Similarly, the correlation (0.5) found between the log values of the ascend and descend time can be explained with the same process: the longer it takes to open the magnetic field up, the longer it takes to close it down.

We found negative correlation (-0.4) between log values of the dimming area and the change in the LOS magnetic field skewness (from pre-eruption to peak-time). This suggests that during their growth phase smaller dimmings acquire a higher magnetic flux imbalance (become more unipolar) than larger dimmings. Hence, the magnetic field reconfiguration and the movement of the loop footpoints is more concentrated in smaller dimmings. We see a similar negative correlation (-0.3) in the relationship between the log values of the dimming area and the change in the mean LOS magnetic field change (from pre-eruption to peak-time), affirming the above explanation: the mean magnetic field becomes stronger in smaller dimmings.

Contrary to previous studies that are based on difference image dimmings, we found no correlation between the dimming area and the CME mass. This is likely because in our study we are observing the footpoints of the erupting loop structure rather than the combination of the footpoints and the expanding CME plasma. We suggest that the area of the footpoints is less indicative of the mass that has been removed. Our results show that the mean peak-time EUV intensity of the dimming negatively correlates with the log of the CME mass (-0.4), kinetic energy (-0.4) and the flare duration (-0.3). Therefore, it is the peak-time intensity (which is related to the local density) that is more indicative of how much plasma was removed and how energetic the eruption was. In other words, the darker the dimming was at its peak, the larger the associated CME mass and kinetic energy was, and the longer the associated flare lasted.

We also found that the pre-eruption mean intensity in the region where the dimming emerges correlates with the log of the CME mass (0.5) and angular width (0.4). Hence, the brighter the region is where the dimming develops, the more mass it can expel and the larger the CME is in its width. A negative correlation (-0.4) was found between the change in the dimming mean intensity in the darkest cores and the CME angular width, the log of the CME mass, and kinetic energy. Similarly, we found that the change in the mean dimming intensity (from pre-eruption to peak-time) also negatively correlates with the log of the CME mass (-0.4) and the CME angular width (-0.4): i.e., the darker the dimming gets, the more mass is removed during the eruption and the larger the CME is in its width. 

When studying flare properties we found that the log of the flare duration negatively correlates with the change in the mean intensity of the whole dimming as well as its darkest core (-0.5 in both cases). This too, supports the reasoning that the more the intensity drops, the larger the eruptive event is.

\subsection{Morphology}

\begin{figure}[!t]
\centerline{
\includegraphics[scale=0.4, trim=0 80 0 0, clip]{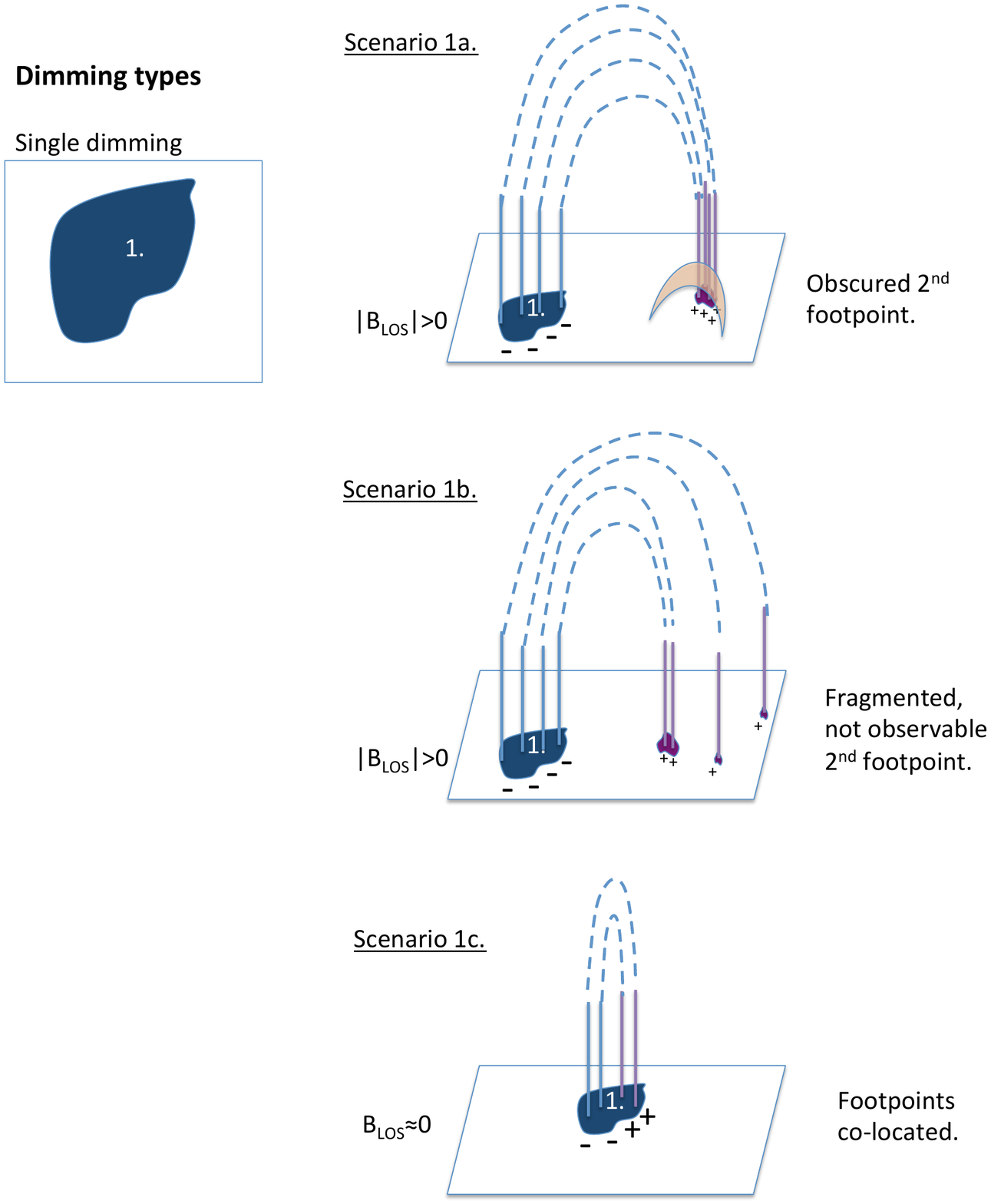}}
\centerline{
\includegraphics[scale=0.4, trim=0 80 0 50, clip]{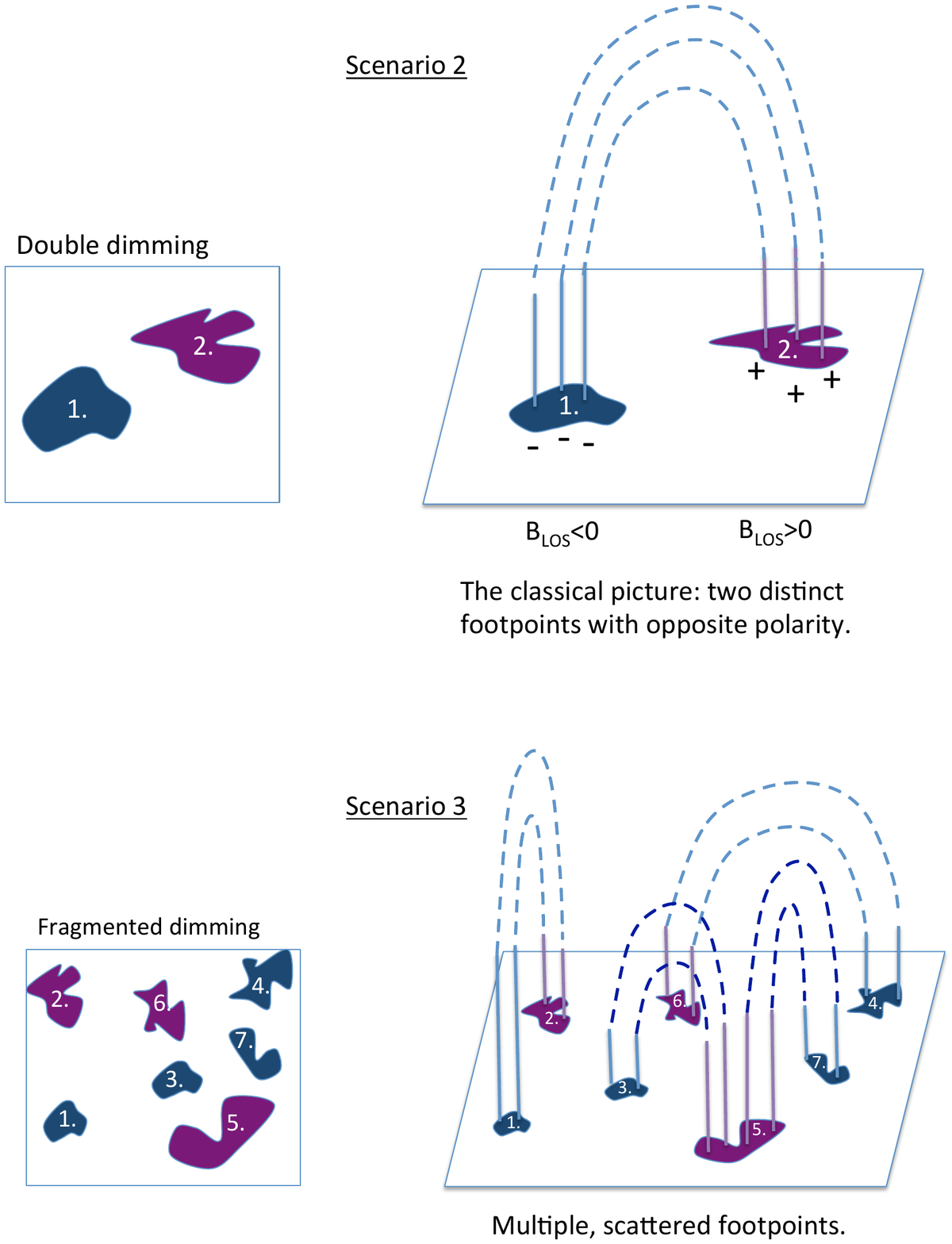}}
\caption{The observed morphological types of dimmings (left) and the magnetic field configuration in each scenario (right). The colours indicate dominant polarity: blue for negative, purple for positive. Scenario 1a: one unipolar footpoint is observed, the other one is obscured. Scenario 1b: one unipolar footpoint observed, the other footpoint is fragmented and not observable. Scenario 1c: the footpoints are co-located and appear as one dimming (no dominant polarity observed). Scenario 2: the classical double dimming configuration with two observed footpoints of opposite dominant polarity. Scenario 3: the fragmented dimming. The quasi-open field lines connect to multiple locations, and the fragments are relatively small, but observable. The footpoints may or may not have dominant polarities.}
\label{blobs}
\end{figure}

The dimmings in our catalogue were separated into three categories based on their morphology: single, double, fragmented. We explain these morphological classes by different configurations of the quasi-open magnetic field. We do this by considering the morphology, flux imbalance and the fact that the magnetic field in transient open magnetic field regions are only ``quasi'' open, and hence any field line must connect back to the Sun at some location. The first class is `single' dimmings, which either have a predominant polarity or not. In order to explain the morphology and flux distribution, we suggest three scenarios (Figure \ref{blobs}). In the first scenario there is a single large dimming with a predominant polarity - this observed dimming is one of the footpoints of the loop structure that erupted. The second dimming (the opposite polarity footpoint) is not observed. This is either because the second footpoint is obscured (Scenario 1a) by bright phenomena (e.g., the AR or post-eruptive loop structure) that are associated with the eruption site, or the second footpoint is simply not concentrated enough into a footpoint or footpoints to be observable and instead is fragmented into multiple footpoints that are challenging to detect (Scenario 1b). In Scenario 1c there is a single dimming without a predominant polarity. In this case, as there is approximately the same amount of positive and negative flux in the same open magnetic field region - we suggest that the footpoints of the erupted loop structure in this case are co-spatial -- meaning that the footpoints are close enough to appear as one large open magnetic field region. Scenario 2 is the `double' dimming morphology -- the well-known configuration, where the footpoints of the loop structure appear as two distinct dimmings with opposite predominant polarities. Incidentally, this ``perfect'' scenario is less common. Scenario 3 is the `fragmented' dimming configuration, where we observe multiple dimmings -- the footpoints of multiple erupted loops. These may or may not have predominant polarities, depending on the spatial distribution of the footpoints. This scenario is the most complex and may also involve one or more of the previous scenarios.

The most prevalent type of dimming in our database is the single dimming (85 detected). The second most common type is the double dimming configuration (22 detected). And the fragmented configuration was the least common (8 detected).

The primary goal of our work was to investigate the association between dimming, flare and CME properties to help us understand the relationship between the different parts of this complex eruptive process: the eruption, the open field created through the removal of plasma and extension of magnetic fields into the corona and beyond. We were particularly interested in the relationship between dimmings and CMEs, since using dimmings could help forecast large geomagnetic storms. In our current catalogue we were able to link 64 dimmings with CMEs. We plan to extend our catalogue for the whole SDO period to include more data and also separate dimmings into different categories (AR-related and QS dimmings) and associate the corresponding geomagnetic disturbance indices (Kp) to extend our investigation. Separating dimmings into different categories might show us whether there is a difference in their evolution and a connection with their eruptive counterparts. To carry out this work, we plan to fully automate CoDiT.

The next stage of our work is to study the observational manifestation of interchange reconnection -- a process that might be responsible for the evolution of the erupted loop structure footpoints. For this we will investigate the LOS magnetic field in the region where the dimming emerges. Initial results suggest that the quasi-open magnetic field might be created through different processes. We have studied a few dimmings and found that the skewness of the LOS magnetic field may or may not change over the lifetime of the dimming. In some cases the skewness is constant, suggesting that the magnetic field opens into an area with an already predominant polarity. In other cases the skewness changes, which suggests that interchange reconnection might be responsible for the displacement of the loop footpoints into the region where the dimming emerges. Our publication on this subject is forthcoming.

\acknowledgments
We would like to thank Dr. Scott McIntosh, Dr. Barbara Thompson, and Dr. Len Fisk for insightful discussions, and the AIA and HMI science teams for the SDO/AIA and HMI data. Dr. Krista would also like to thank NOAA/SWPC and NCAR/HAO for hosting her for many years as a scientist and a long-term visitor, respectively. This material is based upon work supported by the National Aeronautics and Space Administration under grant No. NNX15AB91G issued through the NASA/LWS Program. This work was in part supported by NASA Program Element NNH14ZDA001N-HSR.

{\it Facilities:} \facility{SDO (AIA \& HMI), 
SOHO (LASCO)}

\bibliographystyle{apj} 
\bibliography{biblio}  

\end{document}